\documentstyle[pre,aps,epsf,multicol]{revtex}
\begin{document}

\title{Phase transitions of the binary production $2A\to 3A$, 
$4A\to\emptyset$ model}
\author{G\'eza \'Odor}
\address{Research Institute for Technical Physics and Materials Science, \\
H-1525 Budapest, P.O.Box 49, Hungary}    
\maketitle

\begin{abstract}
Phase transitions of the $2A\to 3A$, $4A\to\emptyset$ 
reaction-diffusion model is explored by dynamical, $N$-cluster 
approximations and by simulations.
The model exhibits site occupation restriction and explicit diffusion
of isolated particles. While the site mean-field approximation shows a
single transition at zero branching rate introduced in
PRE {\bf 67}, 056114 (2003), $N>2$ cluster approximations
predict the appearance of another transition line for weak diffusion ($D$)
as well. The latter phase transition is continuous, occurs
at finite branching rate and exhibits different scaling behavior.
I show that the universal behavior of these transitions is in agreement with 
that of the PCPD model both on the mean-field level and in one dimension.
Therefore this model exhibiting annihilation by quadruplets 
does not fit in the recently suggested classification of universality 
classes of absorbing state transitions in one dimension
(PRL {\bf 90}, 125701 (2003)). For high diffusion rates the effective
$2A\to 3A\to 4A\to\emptyset$ reaction becomes irrelevant and the model
exhibits a mean-field transition only. The two regions are separated by 
a non-trivial critical endpoint at $D^*$.
\end{abstract}

\begin{multicols}{2}

Phase transitions in nonequilibrium systems, which do not possess
hermitian Hamiltonian may appear in models of population, epidemics, 
catalysis, cooperative transport \cite{DickMar}, enzyme biology 
\cite{Berry} and markets for example \cite{Bou}.
Reaction-diffusion systems are of primary interest since other 
nonequilibrium models often can be mapped onto them \cite{Hin2000}.
The classification of universality classes in reaction-diffusion
systems \cite{dok,Uweof} has recently got some impetus. In these systems
particle creation, annihilation and diffusion processes compete and
by tuning the control parameters phase transition may occur from an active
steady state to an inactive, absorbing state of zero density. 
The fluctuations in
the absorbing state are so small, that systems cannot escape from it,
hence such phase transitions may emerge in one dimension already.
Several systems with binary, triplet or quadruplet, particle reactions 
have been investigated numerically and unclassified type of
critical phase transitions were found 
\cite{GrasBP,Carlon99,Hayepcpd,Odo00,HayeDP-ARW,coagcikk,NP01,MHUS,HH,binary,OSC02,multipcpd,PK66,pcpd2cikk,KC0208497,DM0207720,PHK02,tripcikk,brazcikk}.
Solid field theoretical treatment exists for bosonic, binary production
systems only \cite{HT97},
but this is not applicable for the active and critical states of site 
restricted models, since it cannot describe a steady state with 
finite density. 

The mean-field solution of general,
\begin{equation}
n A \stackrel{\sigma}{\to} (n+k)A,
\qquad m A \stackrel{\lambda}{\to} (m-l) A, \label{genreactions}
\end{equation}
models (with $n>1$, $m>1$, $k>0$, $l>0$ and $m-l\ge 0$) 
resulted in a series of universality classes depending on $n$ and $m$
\cite{tripcikk}.
In particular for the $n=m$ symmetrical case the density of particles
above the critical point ($\sigma_c > 0$) scales as
\begin{equation}
\rho \propto |\sigma-\sigma_c|^{\beta}, \label{betascale}
\end{equation}
with $\beta^{MF}=1$, while at the critical point it decays as
\begin{equation}
\rho \propto t^{-\alpha} \ , \label{alphascale}
\end{equation}
with $\alpha^{MF}=\beta^{MF}/\nu^{MF}_{||}=1/n$ \cite{PHK02,tripcikk}
(here "MF" denotes mean-field value).
On the other hand for the $n<m$ asymmetric case continuous phase 
transitions at zero branching rate $\sigma_c=0$ occur with
\begin{equation}
\beta^{MF}=1/(m-n), \quad \alpha^{MF}=1/(m-1) \label{asymmfscal}
\end{equation}
For $n>m$ the mean-field solution provides first order
transition.

The upper critical dimension for such systems is debated
\cite{Uweof,KC0208497,PHK02,tripcikk} but should be quite low
($d_c=1-2$) allowing a few anomalous critical transitions only.
For example $d_c < 1$ was confirmed by simulations in case of the
asymmetric, binary production $2A\to 4A$, $4A\to 2A$ model \cite{brazcikk}.
It was also pointed out there that $N>1$ cluster mean-field 
approximation, that takes into account the diffusion of particles 
would provide a more adequate description of such models. Earlier studies 
have shown \cite{boccikk,OdSzo,meorcikk} that there exist models
with first order transitions in the site mean-field approximation that
changes to continuous one in higher level of cluster approximations.
Dependence on the diffusion was found to be important in binary
production models \cite{Carlon99,Odo00} and it turned out 
that at least $N>2$ level of approximation is needed 
for an adequate description \cite{OSC02,pcpd2cikk}. 

In this paper I investigate the $2A\to 3A$, $4A\to\emptyset$ model and 
show that the diffusion plays an important role: 
it introduces a different critical 
point besides the one at $\sigma=0$ branching rate. I show by simulations that
this transition is not mean-field type in one dimension but belongs to the 
class of the: $2A\to 3A$, $2A\to\emptyset$ so called PCPD model. 
My model is defined and parametrized following
the notation of \cite{Carlon99} by the rules
\begin{eqnarray}
AA\emptyset,\,\emptyset AA \rightarrow AAA  \qquad {\rm with \ rate}
\, & \ \ \sigma = (1-p)(1-D)/2 \nonumber \\
AAAA \rightarrow \emptyset\emptyset\emptyset\emptyset \qquad  
{\rm with \ rate}\,  & \lambda = p(1-D) \nonumber \\
A\emptyset \leftrightarrow \emptyset A \qquad {\rm with \ rate}\,  & D \ .
\label{DynamicRules}
\end{eqnarray}
Here $D$ denotes the diffusion probability and $p$ is the other control
parameter of the system.

Dynamical cluster mean-field approximations have been introduced for
nonequilibrium models by \cite{gut87,dic88}. The master equations for
$N=1,2,3,4,5$ block probabilities were set up
\begin{equation}
\frac{\partial P_N(\{s_i\})}{\partial t} = f\left (P_N(\{s_i\})\right) \ ,
\label{mastereq}
\end{equation}
where site variables may take values: $s_i=\emptyset,A$. 
Taking into account spatial reflection symmetries of $P_N(\{s_i\})$ this 
involves 20 independent variables in case of $N=5$.
The master equation (\ref{mastereq}) was solved numerically using the 
Runge-Kutta algorithm for $N=2,3,4,5$ by several $D$ and $p$ values.
The particle ($\rho(p,D)$) and pair ($\rho_2(p,D)$) densities were determined
by $P_N(\{s_i\})$. 
For strong diffusion rates only a mean-field phase 
transition occurs at $\sigma=0$ with $\beta=1/2$ and $\alpha=1/3$ exponents
belonging to the set set of classes (\ref{asymmfscal}) 
discovered in \cite{tripcikk}.

\begin{figure}
\begin{center}
\epsfxsize=70mm
\centerline{\epsffile{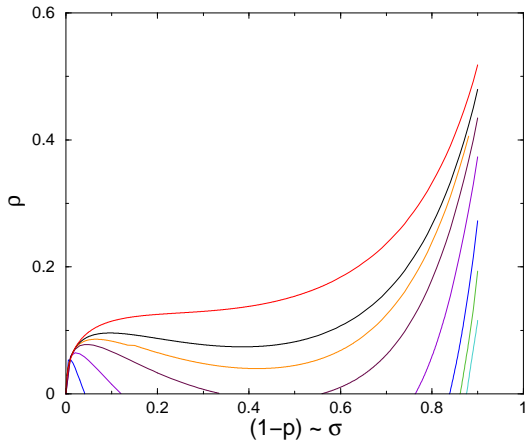}}
\caption{Steady state density in $N=5$ level approximation for diffusion
rates $D=0.5$, $0.4$, $0.35$, $0.3$, $0.2$, $0.1$, $0.05$, $0.01$, 
$0.01$ (top to bottom curves). Similar scenario appears for $N=2,3,4$.}
\label{5GMFpd}
\end{center}
\end{figure}

However for $N>1$ and weak diffusion rates other phase transitions points
emerge as well, with $\sigma_c >0$.
This means that for intermediate $\sigma$ and small $D$ values the 
absorbing state becomes stable as one can see in Fig.\ref{5GMFpd}.
Simulations in one dimension confirm this (see later).
In the active phases in the neighborhood of the $\sigma_c > 0$
transition points power law fitting of the form (\ref{betascale})
to the mean-field data point
resulted in $\beta=1$ for all $N>1$ levels of approximations.
On the other hand for the pair density in pair approximations
one obtains $\beta=1$ again, like in case of the PCPD model for low 
diffusions \cite{Carlon99}. This anomaly disappears 
for $N=3,4,5$ and the fitting results in $\beta=2$ for pairs.

At the $\sigma_c > 0$ critical points the dynamical behavior is
power-law type (\ref{alphascale}) with $\alpha=1/2$ for $N=3,4,5$.
Again the pair approximation gives the strange result: 
$\alpha=1$ (like in ref.\cite{Carlon99}). The failure of the pair
approximation also appears in the inactive region, where it results
in exponential density decay.  In contrast with this the $N=3,4,5$ 
approximations show power-laws here with $\alpha=1$ for particles
and $\alpha=2$ for pairs.
The above $\alpha$ and $\beta$ exponents occurring by low diffusions 
at the critical points and inactive phases in the phase diagram away 
from the $\sigma=0$ transition are different from those
of the site mean-field values (\ref{asymmfscal}). This can be explained 
by accepting that the dominant decay process for $\sigma>0$, $D<D^*$ is:
$2A\to\emptyset$ (via $2A\to 3A\to 4A\to\emptyset$) instead of the 
$4A\to\emptyset$ -- that is the only mode of decay at $\sigma=0$.
Altogether one can find very similar cluster mean-field behavior as 
in case of the PCPD model \cite{Carlon99,pcpd2cikk}.

One can also observe that by increasing $D$ from zero the PCPD
like transitions disappear at some $D^*$ value,
when the $\rho(\infty)$ steady state curve touches the $\rho=0$ axis.
For $D \ge D^*$ there is no absorbing state in the system and a
critical endpoint appears with $\beta=2$ (parabolic) singularity
at $\sigma_c^*$. For $N=5$ the endpoint is located at $D^*=0.301(1)$,
$p_c^*=0.53(1)$.

To test these analytical findings I have performed simulations in one 
dimension.
These were carried out on $L=1-5\times 10^5$ sized systems with periodic 
boundary 
conditions. The initial states were half filled lattices with randomly
distributed $A$-s and the density of particles is followed up to
$5\times 10^8$ Monte Carlo steps (MCS). 
One MCS consist of the following processes. 
A particle, a direction and a number $x \in (0,1)$ are 
selected randomly; if $x < D$ a site exchange is attempted with one of 
the randomly selected empty nearest neighbors (nn);
else if $D \le x < (D+\lambda)$ four neighboring particles are removed;
else one particle is created at an empty site in the randomly selected 
direction following a pair of $A$-s. In each MCS the time is updated
by $1/n$, where $n$ is the number of particles.
\begin{figure}
\epsfxsize=70mm
\epsffile{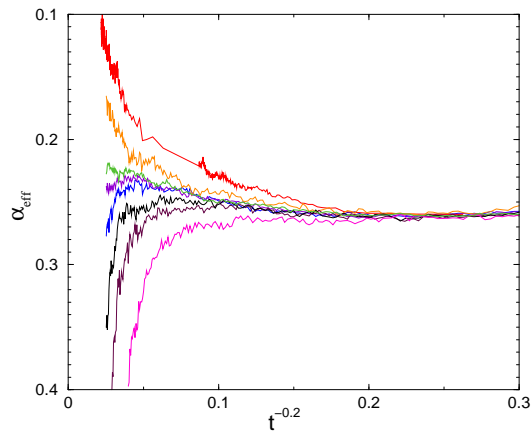}
\caption{$\alpha_{eff}$ in the one dimensional $3A\to 4A$,
$4A\to\emptyset$ model at $D=0.5$
Different curves correspond to $p=0.1583$, $0.1584$, $0.1585$, $0.15852$,
$0.15853$, $0.1586$, $0.1587$, $0.159$ (top to bottom)}
\label{sl5}
\end{figure}

First I followed the density of particles for a small $\sigma$ 
(at $p=0.95$) at diffusion rates: $D=0.5$ and $D=0.2$. 
In both cases a power-law decay with $\alpha = 0.5$ exponent could be 
observed, hence an inactive phase with decay of the $AA\to\emptyset$ 
process --- valid in one dimension \cite{Lee} --- was identified. 

The critical points were determined by calculating the local slopes
defined as
\begin{equation}
\alpha_{eff}(t) = {- \ln \left[ \rho(t) / \rho(t/m) \right] 
\over \ln(m)} \ , \label{slopes}
\end{equation}
(where I used $m=2$) for $D = 0.2, 0.5, 0.747$.
As Fig.\ref{sl5} shows the local slopes curve for $D=0.5$, $p=0.15850(2)$ 
extrapolates to $\alpha=0.21(1)$. This value agrees with that of the PCPD 
model \cite{pcpd2cikk,KC0208497}.
Other curves exhibit curvature for long times,
i.e. for $p < 0.1585$ they veer up (active phase), while for
$p > 0.1585$ they veer down (absorbing phase).
The local slopes figure shows similar strong correction to
scaling as in case of the PCPD model, i.e. some curves that seem to
be supercritical veer down after $t > \sim 10^6$ MCS.
Similar results are obtained by other $\sigma_c > 0$ transitions.
For $D=0.2$, when the critical point is at $p=0.0892(1)$ the local slopes 
for the density decay predicts $\alpha=0.21(2)$. 
Repeating the simulations at $D=0.9$ no absorbing phase
has been found (up to $p\le 0.9999$), the steady state density disappears 
monotonously as $\sigma\to 0$. At $\sigma=0$ the density decays with 
$\alpha=1/3$ valid for the $4A\to\emptyset$ process in one dimension 
\cite{Cardy-Tauber}.

\begin{figure}
\begin{center}
\epsfxsize=70mm
\centerline{\epsffile{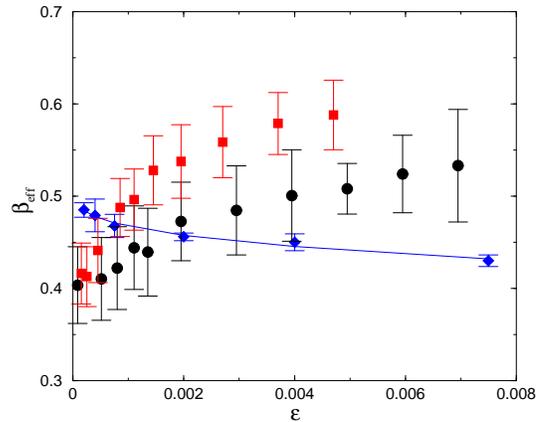}}
\caption{$\beta_{eff}$ as the function of $\epsilon$ in the
one dimensional $2A\to 3A$, $4A\to\emptyset$ model.
The bullets correspond to $D=0.5$, the boxes to $D=0.2$, the diamonds
to $D=0.9$ diffusion rate. The solid line shows a quadratic fitting
of the form (\ref{fitform}).}
\label{beta}
\end{center}
\end{figure}

The steady state density in the active phase near the critical
phase transition point is expected to scale as 
$\rho(\infty)\propto|p-p_c|^{\beta}$.
Using the local slopes method one can get a precise estimate for
$\beta$ and see the corrections to scaling
\begin{equation}
\beta_{eff}(p_i) = \frac {\ln \rho(\infty,p_i) -
\ln \rho(\infty,p_{i-1})} {\ln(p_i) - \ln(p_{i-1})} \ .
\label{beff}
\end{equation}
The steady state density was determined by running the simulations
in the active phase: $\epsilon=p_c-p_i>0$, by averaging over $\sim 100$ 
samples in a time window following the level-off is achieved.
As one can see on Fig.\ref{beta} the effective exponent tends
to $\beta=0.40(2)$ as $\epsilon\to\emptyset$ both for $D=0.5$ and 
$D=0.2$ diffusions. 
These values are in agreement with that of the one dimensional
PCPD model \cite{pcpd2cikk,KC0208497}.
Again assuming logarithmic corrections as in \cite{pcpd2cikk} of the form
\begin{equation}
\rho(\infty,\epsilon) = \left[ \epsilon/ (a+ b\ln(\epsilon))\right]^{\beta} 
\label{logcor}
\end{equation}
one can obtain $p=0.1585(1)$ and $\beta=0.38(1)$ for $D=0.5$ and
$p=0.0892(1)$ and $\beta=0.41(3)$ for $D=0.2$, which agrees with
the previous values within numerical accuracy.
\cite{HayeDP-ARW,tripcikk}.
Altogether one can not see relevant logarithmic corrections for the
diffusion rates investigated here.

\begin{figure}
\begin{center}
\epsfxsize=70mm
\centerline{\epsffile{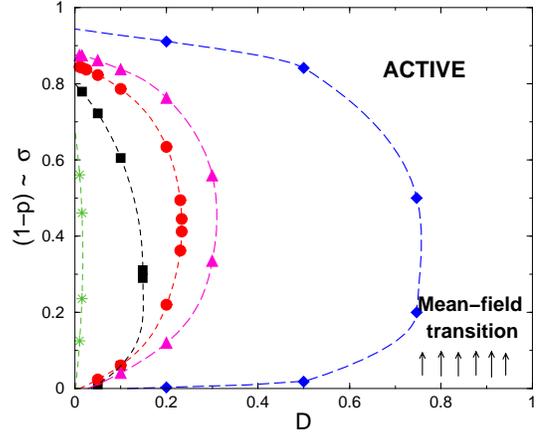}}
\caption{Phase diagram. Stars correspond to $N=2$, boxes to $N=3$,
bullets to $N=4$ and triangles to $N=5$ cluster mean-field approximations. 
Diamonds denote 1d simulation data. The lines serve to guide the eye. 
At the $\sigma=0$ line a mean-field transition occurs.}
\label{pd}
\end{center}
\end{figure}

In case of $D=0.9$, $\sigma_c=0$ one can see $\beta=0.50(1)$ in agreement 
with the $N=3,4,5$ cluster mean-field approximation results. 
A quadratic fitting of the form
\begin{equation}
\beta_{eff} = \beta-a\epsilon^{x}-b\epsilon^{2x} \label{fitform}
\end{equation}
results in: $a=0.195$, $b=0.158$, $x=0.214$, $\beta=0.51(1)$.
This suggests that the effective $2A\to\emptyset$
process is weaker now than the $2A\to 3A$, leaving the transition at 
$\sigma_c=0$.
The phase diagram for different levels of approximations as well as MC data
are shown on Fig. \ref{pd}. As one can see approximations tend towards the
simulated points by increasing $N$. 
Similar reentrant phase diagram has been observed in case of the unary 
production, triplet annihilation model ($A\to 2A$, $3A\to\emptyset$)
\cite{Dicktrip} and in a variant of the NEKIM model \cite{nbarw2cikk}.
In all cases the diffusion competes with particle reaction processes,
and the bare parameters should somehow form renormalized reaction rates
which govern the evolution over long times and distances, the details have
not been worked out.

Finite size scaling investigations at $D=0.5$ and $p_c=0.1585$ were 
performed for system sizes: $L_i=32,64,128...4096$. 
The quasi-steady state density 
(averaged over surviving samples) is expected to scale according to
\begin{equation}
\rho_s(\infty,p_c,L) \propto L^{-\beta/\nu_{\perp}} , \label{betapnfss}
\end{equation}
while the characteristic lifetime for half of the samples to reach the
absorbing state scales with the dynamical exponent $Z$ as
\begin{equation}
\tau(p_c,L) \propto L^Z \ .
\end{equation}
These quantities were analyzed by the local slopes:
\begin{eqnarray}
Z_{eff} (L) = \frac {\ln \tau(L_i) -\ln \tau(L_{i-1})}
              {\ln L_i - \ln L_{i-1}} \\
\beta/\nu_{\perp} (L) = \frac {\ln \rho_s(L_i) -\ln \rho_s(L_{i-1})}
                   {\ln L_i - \ln L_{i-1}} \ \ ,
\end{eqnarray}
Linear extrapolation to $L\to\infty$ results in $Z=1.80(15)$ and 
$\beta/\nu_{\perp}=0.40(3)$. 
These values corroborate that the transition is of PCPD type

In conclusion the $N$ cluster mean-field study of the 
binary production $2A\to 3A$, $4A\to\emptyset$ model has shown the
appearance of another critical transition with non-zero production
rate for low diffusions.
While the pair approximation results in somewhat odd results
-- like in case of other binary production systems --
the $N=3,4,5$ levels coherently exhibit PCPD-like mean-field
critical behavior for these phase transition points and within the absorbing 
phase.
This transition line disappears at a critical endpoint for $D \ge D^*$
characterized by $\beta=2$ order parameter singularity and for high diffusion 
rates the $\sigma_c=0$ critical point remains only in the system, predicted by 
the site mean-field approximation. The utmost importance of diffusion 
dependence and the corresponding $N>2$ cluster mean-field approximations is
demonstrated in this study. 

Extensive simulations in one dimension have confirmed the existence of
the nontrivial transition for low diffusions. By these transitions points
the critical behavior agrees with that of the latest results obtained for 
the PCPD model.
Therefore this model does not fit in the table of universality classes 
suggested for such models in one dimension \cite{KC0208497}. 
The reason behind this discrepancy might be that in \cite{KC0208497} low 
diffusions have not been investigated or the lack of complete site exclusion
in their model. Site exclusion has been shown to be relevant in multi-species  
reaction-diffusion systems and in binary production systems \cite{OdMe02}.

An interesting, open problem is the exploration of the phase structure 
of this system in higher dimensions. 
The agreement of one-dimensional results with those of the cluster 
mean-field shows that similar rich phase structure may emerge in higher
dimensions too. That would mean that the effective $2A\to\emptyset$ reaction 
is generated via: $2A\to 3A\to 4A\to\emptyset$ again.
These results raise the possibility that such mechanism also emerges by unary 
production systems (example by: $A\to 2A$, $4A\to\emptyset$) and one should
find a DP transition instead of the mean-field one suggested by perturbative 
renormalization study \cite{Cardy-Tauber} of such models. This would affect 
the classification of fundamental universality classes of RD systems and
may point out the weak points of the perturbative renormalization.
An other important point to be investigated is the scaling behavior
at the critical endpoint.
\bigskip

{\bf Acknowledgements:}\\

The author thanks F. Igl\'oi, G. Szab\'o and U. T\"auber for their
useful comments.
Support from the Hungarian research fund OTKA (Grant No. T-25286)
is acknowledged. The author thanks the access to the NIIFI Cluster-GRID 
and to the Supercomputer Center of Hungary.

\end{multicols}
\end{document}